\documentclass[11pt,a4paper,leqno]{article}
\usepackage{graphicx}
\usepackage{a4wide}
\usepackage{amsmath}
\usepackage{color}
\graphicspath{{figures/}}

\begin{document}

\title{Sizing of Hall effect thrusters with input power and thrust level: An Empirical Approach}
\author{K. Dannenmayer$^\dagger$, S. Mazouffre$^\dagger$\footnote{Author to whom correspondence should be addressed: stephane.mazouffre@cnrs-orleans.fr}}
\maketitle
$^\dagger$ {\small ICARE, CNRS, 1C avenue de la Recherche Scientifique, 45071 Orl\'{e}ans, France.}\\

\begin{abstract}
Sizing methods can be used to get a first estimate of the required
Hall thruster dimensions and operating conditions for a given input
power and a corresponding thrust level. After a review of the
existing methods, a new approach, which considers the three
characteristic thruster dimensions, i.e. the channel length, the
channel width and the channel mean diameter as well as the magnetic
field, is introduced. This approach is based on analytical laws
deduced from the physical principles that govern the properties of a
Hall effect thruster, relying on a list of simplifying assumptions.
In addition, constraints on the channel wall temperature as well as
on the propellant atom density inside the channel are taken into
account. The validity of the scaling laws is discussed in light of a
vast database that comprises 23 single-stage Hall effect thrusters
covering a power range from 10\,W to 50\,kW. Finally, the sizing
method is employed to obtain a preliminary geometry and the magnetic
field strength for a 20\,kW and a 25\,kW Hall effect thruster able
to deliver a thrust of 1\,N, respectively 1.5\,N.
\end{abstract}
\mbox{}\newline
\newline
\noindent To be published in {\it Journal of Technical Physics} {\bf
49}, vol. 3-4 (2008).

\newpage
\section*{Nomenclature}
\begin{tabbing}
$T_{ext}, T_{max}, T_{int}$ \quad\= temperature \kill
$A$ \> channel cross section; area\\
$B$ \> magnetic field strength\\
$C_{I_{sp}}$ \> analytic proportionality coefficient for $I_{sp}$\\
$C_{I_{sp}}^*$ \> empiric proportionality coefficient for $I_{sp}$\\
$C_{T_{1,2}}$ \> analytic proportionality coefficient for $T$\\
$C_{T_{1,2}}^*$ \> empiric proportionality coefficient for $T$\\
$C_P$ \> analytic proportionality coefficient for $P$\\
$C_P^*$ \> empiric proportionality coefficient for $P$\\
$d$, $d_{ext}$, $d_{int}$ \> mean, external and internal channel diameter\\
$G_{ij}$ \> grey body configuration factor\\
$h$ \> channel width\\
$I_d, I_i$ \> discharge current, ion current\\
$I_m$ \> current corresponding to the propellant mass flow rate\\
$I_{sp}$ \> specific impulse\\
$k_B$ \> Boltzmann constant\\
$L$ \> channel length\\
$m_n, m_i, m_e$ \> propellant atom mass, ion mass, electron mass\\
$\dot{m}, \dot{m_i}$ \> propellant mass flow rate through the anode, ion mass flow rate\\
$n_n, n_e, n_i$ \> atom, electron and ion number density\\
$P$ \> input power\\
$P_{loss}$ \> power losses due to plasma-wall interactions\\
$q_p$ \> heat flux deposited by the plasma onto channel walls\\
$r$, $R$ \> inner, respectively outer, channel
radius\\
$r_{Le}, r_{Li}$ \> electron and ion Larmor radius\\
$T$ \> thrust\\
$T_{max}, T_{ext}, T_{int}$ \> maximum, external and internal wall temperature\\
$T_e$ \> electron temperature\\
$U_d$ \> discharge voltage\\
$v_n, v_e, v_i$ \> thermal velocity of atoms, electrons and ions\\
$\bar{v_i}$ \> average ion exhaust velocity\\
$\alpha$ \> propellant conversion efficiency\\
$\beta$ \> ratio between ionization mean free path and channel length\\
$\Delta$ \> voltage losses\\
$\epsilon$ \> emissivity\\
$\eta$ \> thrust efficiency\\
$\lambda_i, \lambda_{ee}$ \> ionization and electron-electron collision mean free path\\
$\nu_{ce}$ \> electron gyrofrequency\\
$\nu_{en}$ \> electron-atom collision frequency\\
$\xi$ \> scaling index variable\\
$\sigma_i, \sigma_{ee}$ \> cross-section for ionization and electron-electron impact\\
$\sigma_{en}$ \> electron-atom momentum exchange cross-section\\
$\tau_c$ \> gyroperiod\\
$\tau_{en}$ \> electron-atom collisional time\\
\end{tabbing}

\newpage
\numberwithin{equation}{section}

\section{Introduction}

Electric propulsion is nowadays a well-established technology for
space applications~\cite{JPL}. Among all proposed electric
propulsive devices such as arcjet, magnetoplasmadynamic thruster,
gridded ion engine and Hall Effect Thruster (HET), the latter is
currently recognized as an attractive propulsion means for long
duration missions and for maneuvers that require a large velocity
increment. Hall effect thrusters, also called Stationary Plasma
Thrusters or closed electron drift thrusters, are advanced
propulsion devices that use an electric discharge with magnetized
electrons to ionize and accelerate a propellant
gas~\cite{Zhurin,GASCON}. Due to interesting features in terms of
propellant ejection speed, efficiency, flexibility and lifetime,
HETs are now employed for missions like geo-stationary satellite
orbit correction and station keeping. Moreover, HETs appear as good
candidates to be used as the primary propulsion engine for space
probes during interplanetary journeys, as demonstrated by the
successful SMART-1 Moon flyby solar-powered mission of the European
Space Agency~\cite{SMART-1}.

\begin{figure}[t]
\centering
\includegraphics[width=6cm]{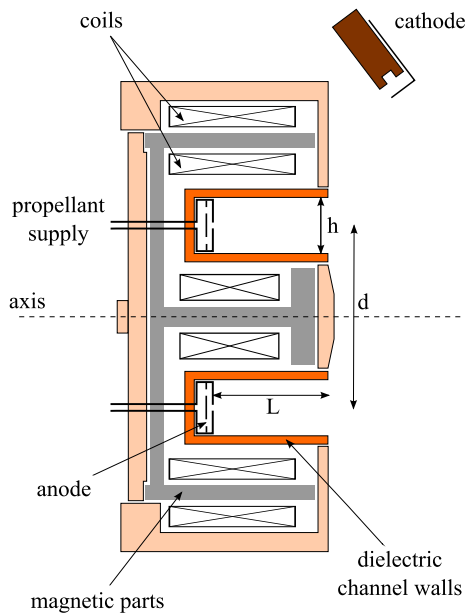}
\caption{Schematic of a Hall effect thruster. The three
characteristic dimensions $L$, $h$ and $d$ are also shown.}
\label{fig:drawing}
\end{figure}

A schematic of a Hall effect thruster is depicted in
Fig.~\ref{fig:drawing}. The basic physics of a HET implies a
magnetic barrier and a low pressure DC discharge generated between
an external hollow cathode and an anode in such a way that a crossed
electric and magnetic fields discharge is
created~\cite{Zhurin,Morozov,Kim}. The anode, which also serves as
gas injector, is located at the upstream end of a coaxial annular
dielectric channel that confines the discharge. Xenon is generally
used as a propellant gas for its specific properties in terms of
high atomic mass and low ionization energy. A set of solenoids
provides a radially directed magnetic field {\bf B} of which the
strength is maximum in the vicinity of the channel exhaust. The
magnetic field is chosen strong enough to make the electron Larmor
radius much smaller than the discharge chamber length, but weak
enough not to affect ion trajectories. The electric potential drop
is mostly concentrated in the final section of the channel owing to
the low electron axial mobility in this restricted area. The
corresponding induced local axial electric field {\bf E} has two
main effects. First, it drives a high electron azimuthal drift --
the Hall current -- that is responsible for the efficient ionization
of the supplied gas. Second, it accelerates ions out of the channel,
which generates thrust. The ion beam is neutralized by a fraction of
electrons emitted from the hollow cathode. When operating near
1.5~kW, a HET ejects ions at 20\,km\,s$^{-1}$ and generates 100~mN
of thrust with an overall efficiency of about 50~\%.

New fields of application are nowadays envisaged for electric
propulsion systems that require low and high power devices. Low
power Hall thrusters ($\sim$\,100 W) are well suited for drag
compensation of observation satellites that operate on a
low-altitude orbit in the Earth atmosphere as well as for orbit
correction and orbit transfer of small platforms. The use of high
power ($\sim$\,5 kW) Hall thrusters for orbit raising and orbit
topping maneuvers of communication satellites would offer
significant benefits in terms of launch mass, payload mass and
operational life. In addition, journeys towards far-off planets and
asteroids with large and heavy robotic probes necessitate to build
thrusters with an input power in the range 10-100\,kW. In view of
the projects demand, it appears necessary to expand the operating
envelope of existing Hall effect thruster technology to achieve the
required performance level. A non-trivial question then arises: How
to extrapolate the design and architecture of currently existing
Hall thrusters towards different scales and different operating
conditions? In other words, what are the scaling laws that connect
Hall effect thruster characteristic dimensions with operating
parameters like discharge voltage, propellant mass flow rate and
magnetic field strength and performances in terms of thrust,
specific impulse and overall efficiency?

Scaling laws that govern the physical properties, the accelerating
ability as well as propellant and energy consumption of Hall
thrusters have been extensively investigated by numerous authors
since the period of development of Hall thrusters in the 70's. In
spite of decades of research on this subject, the assessment of
scaling laws is still a topic of great interest with debates and
controversies as various methodologies and results exist. Therefore,
before describing the approach associated with this study as well as
its outcomes, it is worth briefly reviewing works carried out on
this topic during the past few years and available in the
literature. According to the pioneer works of the Russian physicist
A.~Morozov~\cite{Morozov}, in order to derive scaling laws, it is
necessary to find a similarity criterion, or a set of criteria, that
warrant the underlying physical processes stay unchanged whatever
the truster. This principle states that the properties of thrusters
with a different geometry are linked by way of scaling laws only if
a sufficiently large number of dimensionless similarity criteria
coincide. The complete set of similarity criteria has not yet been
found, however, all works show that the \emph{Melikov-Morozov
criterion} has a strong impact on HET behavior and characteristics
and it must always be taken into account. This criterion indicates
that $\lambda_i \ll L$. In addition to similarity criteria, the
investigation of scaling laws for Hall thrusters necessitates to
account for simplifying assumptions, e.g. a frozen electron
temperature, and constraints like a high efficiency and a reasonable
channel wall temperature.

V.~Zhurin and co-workers propose a size scaling method which is
limited to the effects of changing either the channel width $h$ or
the channel mean diameter $d$~\cite{Zhurin}. One assumption is that
the channel mean diameter is much larger than the channel width, so
that variation of parameters in the radial direction are negligible.
Furthermore, they considered a fixed discharge voltage.
In order to obtain similar performances for two different thruster
configurations, authors shown that the ratio of $r_L$, $\lambda$ and
$L$ to $h$ must stay the same for the two configurations. Using
these criteria and a set of assumptions they demonstrate that the
magnetic field strength is inversely proportional to the channel
width, $B \propto 1/h$, whereas the discharge current and the
propellant mass flow rate are inversely proportional to the channel
mean diameter, $I_d \propto 1/d$ and $\dot{m} \propto 1/d$.

The method presented by J.~Ashkenazy et al. concerns low power
scaling and it is based on the idea of a sufficient propellant
utilization~\cite{Ashkenazy}. They show by means of a simplified
analysis that a straightforward approach for scaling down the
channel size results in a rise of power losses and a reduced overall
thruster efficiency. To avoid these effects, the acceleration region
has to be scaled down along with the channel width and the magnetic
field strength must be increased in proportion to $1/h$. They also
describe an alternative view that consists in extending the channel
length.

M.~Martinez-Sanchez et al. propose an approach for low-power Hall
effect thrusters that includes the use of reference points
\cite{Martinez-Sanchez}. The goal is to achieve a reduction in the
thrusters length scale while preserving both the thrust efficiency
and the specific impulse. The main result of their study is that the
propellant mass flow rate and the applied power scale as the channel
length, $\dot{m} \propto L$ and $P \propto L$, whereas the magnetic
field strength is in inverse proportion to the latter, $B \propto
1/L$. This scaling method allows to calculate the channel size and
the performances of a small thruster with a given input power
provided that a well-known thruster is utilized as a reference.

The team led by M.~Andrenucci suggests to employ an analytical model
coupled to an existing Hall thruster database~\cite{Andrenucci}.
This scaling methodology aims to provide design options for high
power thrusters. The idea is to create a vector of fundamental
parameters defining the thruster geometry and its performances. This
vector is composed of three geometric parameters ($L$, $h$ and $d$),
the gas number density in the injection plane and the applied
discharge voltage $U_d$. A scaling matrix derived from the Hall
thruster physical principles allows to obtain a new thruster
characteristics on the basis of a reference thruster.

In a recent article \cite{DarenYu}, Daren Yu et al. present an
improvement of the existing scaling theory by introducing a scaling
index variable. They assume that the electron temperature and the
discharge voltage are constant, that the ratio $\lambda_i/L$ is
constant and that the geometric similarity is given by
$r/R$=constant and $R^{2-\xi}/L$=constant, where $R$ and $r$ are the
channel outer respectively inner radius and $\xi$ represents a
scaling index variable. They show by way of a comparison of
experimental data with numerical outcomes for different values of
$\xi$, that results obtained from their scaling theory agree well
with the experimental data for $\xi$=2. Therefore they deduce that
the number density $n$ is constant, whereas the mass flow rate
$\dot{m}$, the input power $P$, the thrust $T$ and the discharge
current $I_d$ are proportional to the square of the outer channel
radius $R$.

Finally, it was demonstrated by V. Kim \cite{Kim} that to reach a
high level of efficiency it is not only necessary to ionize and
accelerate ions but to accelerate them into the proper direction,
hence the need for an optimized magnetic field topology. In short,
for a HET with an optimized magnetic field map, there is a
relationship between the acceleration layer length $L_a$ and the
magnetic layer characteristic size $L_B$ and between $L_a$ and $h$.
The use of these similarity criteria, which include the magnetic
field structure, permits to simplify the development of new Hall
thrusters models. V.~Kim et al also emphasize the fact that in the
case of a traditional HET design, the ratio of the ionization mean
free path to the channel length must stay constant ({\it
Melikov-Morozov criterion})~\cite{smallSPT}. Moreover they give
additional criteria about the geometry of thruster elements, i.e.
$L$ and $d$ are both proportional to $h$, that are equivalent to the
ones given by V.~Zhurin. There are two immediate consequences.
First, the propellant mass flow density and the power density rise
when the thruster size decreases, assuming a constant discharge
voltage. Second, as the magnetic field strength is connected with
the characteristic dimensions, notably the channel width, and with
the operation mode, the magnetic field magnitude must rise when the
size
reduces in order to keep comparable conditions.\\

In this contribution, we propose an original way to extrapolate Hall
thruster geometry towards both the low and high power ranges. The
approach is based on the combination of a set of scaling laws with a
vast body of existing data. Besides, realistic constraints on the
performance level and on the thermal load allow to refine the set of
possible $L$, $h$ and $d$ characteristic scale lengths found for a
given thruster input power. In section 2, similarity criteria and
scaling laws are derived from the physics at work in a HET and,
consequently, relationships between performances, operating
parameters and the characteristic dimensions are established. The
thermal constraint is discussed in section~3 along with modeling of
thruster thermal behavior. A detailed description of the database on
Hall thruster performances and properties is given in section~4.
Besides, recorded data are compared with calculation outcomes. In
section~5, our approach is finally employed to design a 20\,kW as
well as a 25\,kW thruster. A summary and general conclusions follow
in section~6.

\section{Set of governing relations}
\subsection{Similarity criteria and scaling laws}
\label{section-ScalingLaws}
A necessary first step in order to
determine scaling laws for Hall effect thruster does consist in
finding some critical parameters as well as in defining similarity
criteria based on the current knowledge and understanding of the
physics of Hall thrusters. The geometry of a Hall effect thruster is
defined by three characteristic dimensions, see
Fig.~\ref{fig:drawing}, the discharge channel length \emph{L}, the
mean channel diameter $d=\frac{1}{2}(d_{ext}+d_{int})$ and the
channel width \emph{h}, as well as by a set of operating parameters
such as the magnetic field strength \emph{B}, the discharge voltage
\emph{$U_d$} and the propellant mass flow rate \emph{$\dot{m}$}.\\
To simplify the assessment of scaling laws, the following
assumptions have been made throughout the entire paper:
\begin{itemize}
\item[-] the electron temperature is constant and homogeneous whatever
the operating conditions,
\item[-] the propellant gas has a uniform temperature all over the channel, hence a constant propellant
velocity,
\item[-] the potential energy is fully
converted into kinetic energy and all ions experience the whole
potential drop, of which the magnitude is $U_d$,
\item[-] plasma-wall interactions are taken into account through heat load to the channel
walls,
\item[-] the magnetic field is uniform; its value at the exit plane
is solely considered,
\item[-] electron transport across the magnetic
barrier is considered as classical: no anomalous transport is
accounted for within the region of strong magnetic
field~\cite{ANOMALOUS},
\item[-] there are no multiply-charged ions in the plasma,
\item[-] a parallel monokinetic ion beam is produced, i.e. the plasma jet divergence
is null.
\end{itemize}
The first relationship reflecting the impact of the thruster scale
on its performance is the relationship between the discharge channel
length \emph{L} and the ionization mean free path
\emph{$\lambda_i$}. To ensure a sufficient ionization of the gas, it
is necessary to satisfy the Melikov-Morozov criterion:
\begin{equation}
\lambda_i \ll L.
\end{equation}
The ionization frequency that originates from electron-atom impacts
reads:
\begin{equation}
\nu_i = n_n \, \langle \sigma_i(v_e) \, v_e \rangle \approx n_n \,
\sigma_i(T_e) \, \sqrt{\frac{8 k_B T_e}{\pi m_e}}.
\end{equation}
The ionization length, which corresponds to the mean distance an
atom can travel before being ionized, can be formulated as the
product $\nu_i \, v_n$ in a first order approximation. Therefore the
Melikov-Morozov criterion can be expressed as:
\begin{equation}
\lambda_i = \frac{v_n}{n_n \, \sigma_i \, v_e(T_e)} \ll L.
\label{lambdaion}
\end{equation}

Assuming that both the electron temperature \emph{$T_e$} and the
propellant speed $v_n$ are constant and independent of the operation
conditions, implies that \emph{$\sigma_i$} and \emph{$v_e$}, being
only a function of \emph{$T_e$}, remain constant. To keep the ratio
$\lambda_i/L$ constant, equation \ref{lambdaion} leads to:
\begin{equation}
n_n\propto \frac{1}{L}.
\label{n1}
\end{equation}
In the same way, the ratio $\lambda_{ee}/L$ must stay constant,
where \emph{$\lambda_{ee}$} is electron-electron impact mean free
path:
\begin{equation}
\lambda_{ee}=\frac{1}{n_e\sigma_{ee}},
\label{lambdaelec}
\end{equation}
where the cross-section $\sigma_{ee}$ is a function of the Coulomb
logarithm~\cite{NRL}. The equation \ref{lambdaelec} implies that:
\begin{equation}
n_e\propto \frac{1}{L}.
\label{ne}
\end{equation}
The relations \ref{n1} and \ref{ne} for the atom, respectively the
electron, number density are the same than those developed before by
V.~Zhurin and M.~Martinez-Sanchez \cite{Zhurin,Martinez-Sanchez}.
Another relation between $n_n$ and Hall thruster dimensions can be
established when considering the propellant mass flow rate passing
through the anode. The quantity $\dot{m}$ can be decomposed into the
product of several terms:
\begin{equation}~
\dot{m}=n_n\cdot m_n\cdot v_n\cdot A.
\label{debit}
\end{equation}
The annular channel cross section $A$ is given by:
\begin{equation}
A=\frac{\pi}{4}(d_{ext}^2-d_{int}^2)=\frac{\pi}{4}\underbrace{(d_{ext}+d_{int})}_{2d}\underbrace{(d_{ext}-d_{int})}_{2h}=\pi
dh,
\end{equation}
hence $A\propto hd$. We can therefore consider that for a constant
gas mass flow rate:
\begin{equation}
n_n\propto \frac{1}{hd}.
\label{n2}
\end{equation}
This second relationship between the atom number density and the
thruster dimensions $h$ and $d$ has never been mentioned previously,
as authors never considered two sizes at the same time. Note that,
in order to keep the physical processes at work in a Hall effect
thruster unchanged, to warrant a high efficiency and to limit the
thermal load as well as the channel wall wear, the number densities
of electron and atoms must stay roughly fixed inside the thruster
channel whatever the input power. The average values commonly found
in literature, which turn out to guarantee a satisfying operation,
are:
$n_n=10^{19}$\,m$^{-3}$ and $n_e=10^{18}$\,m$^{-3}$.\\
Another interesting fact is that the variables $L$, $h$ an $d$ are
linked to one another. Indeed, by using the Melikov-Morozov
criterion and the mass flow rate definition, it comes:
\begin{eqnarray*}
\frac{\lambda_i}{L}= \beta \;\; \Rightarrow \;\;
\frac{v_n}{n_n\sigma_iv_e}=\beta L \;\; \Rightarrow \;\;
n_n=\frac{1}{L}\cdot \frac{v_n}{\sigma_iv_e\beta}\\
\dot{m}=n_n\cdot m_n \cdot v_n \cdot \pi hd \;\; \Rightarrow \;\;
n_n=\frac{\dot{m}}{m_n v_n \pi}\cdot \frac{1}{hd}\\
\end{eqnarray*}
Combining the two previous relations, one finds:
\begin{equation}
\frac{hd}{L}=\frac{\sigma_i v_e \beta \dot{m}}{m_n v_n^2 \pi} =
f(\beta,\dot{m}).
\end{equation}
Therefore, for a fixed value of the $\beta$ parameter, and under our
list of assumptions, the thruster characteristic dimensions are
coupled through the gas mass flow
rate.\\

\noindent The magnetic field strength in a Hall effect thruster is
such that electrons are magnetized and ions are not magnetized. The
following criterion must then be fulfilled~\cite{Kim}:
\begin{equation}
r_{Le}\ll L\ll r_{Li}.
\end{equation}
The definition of the electron Larmor radius is:
\begin{equation}
r_{Le}=\frac{m_ev_e}{eB}.
\end{equation}
Using the fact that the ratio of $r_{Le}$ to $L$ must remain
constant, the following relationship can be established:
\begin{equation}
B\propto \frac{1}{L}.
\label{B1}
\end{equation}
The relation~\ref{B1} between \emph{B} and \emph{L} has already been
mentioned by M.~Martinez-Sanchez \cite{Martinez-Sanchez}. A second
constraint for the magnetic field strength can be established due to
the fact that the electron gyroperiod $\tau_{ce}$ in the magnetic
barrier must be much shorter than the time between two consecutive
electron-atom collisions:
\begin{equation}
\frac{\tau_{en}}{\tau_{ce}}=\frac{\nu_{ce}}{\nu_{en}}\gg1,
\label{collision}
\end{equation}
where $\nu_{ce}=eB/m_e$ and \(\nu_{en}=n_n\sigma_{en}v_e\)
\cite{Rax}. This strong point indicates that electrons must be
efficiently trapped inside the magnetic field of a Hall thruster in
order to produce a high electric field and to favor ionization of
the seeded gas.. In fact, $\tau_{en}$ is so long in a HET that
anomalous electron transport perpendicular to the magnetic field
lines must be put forwards to correctly explain the observed
properties and the magnitude of measured
quantities~\cite{Zhurin,Morozov}. Assuming that the ratio $\nu_c /
\nu_{en}$ must remain constant, Eq.~\ref{collision} implies:
\begin{equation}
B\propto \frac{1}{hd}.
\label{B2}
\end{equation}
To the best of our knowledge, the correlation between \emph{B} and a
product of two dimensions has never been mentioned before as solely
one characteristic thruster dimension is usually considered by
authors. Combining equations 2.9 and 2.15, one obtains $B \propto
n_n$ in compliance with the fact that plasma containment depends on
collision events with neutrals.

\subsection{Relationship between performances and dimensions}
\label{section_relation}
The previously derived scaling laws can now
be utilized to analyze the effect of dimensions upon the
performances of a Hall effect thruster.

\noindent The specific impulse is defined as follows:
\begin{equation}
I_{sp}=\frac{\bar{v_i}}{g_0},
\end{equation}
where $g_0$ is the standard gravity at Earth's surface. The average
ion exhaust velocity can be defined as~\cite{Morozov}:
\begin{equation}
\bar{v_i}=\sqrt{\frac{2e}{m_i}(U_d-\Delta)}.
\end{equation}
Voltage losses are neglected in this work. Therefore the ion exhaust
velocity is proportional to the square root of the discharge voltage
(\(\bar{v_i}\propto \sqrt{U_d}\)) and for this reason:
\begin{equation}
I_{sp}\propto \sqrt{U_d}.
\label{Isp}
\end{equation}
The thrust of a Hall effect thruster reads:
\begin{equation}
T=\dot{m_i}\bar{v_i}=\alpha\,\dot{m}\,\bar{v_i},
\end{equation}
where the coefficient $\alpha$ stands for the fraction of propellant
atoms that are converted into ions. Considering in a first approach
that $\alpha$ is constant --\:the typical value given in the
literature is $\alpha \approx 0.9$\:-- and using equation
(\ref{debit}), one can establish the following relationship for the
thrust:
\begin{equation}
T\propto \frac{1}{L}\,\sqrt{U_d}\,hd \label{thrust}.
\end{equation}
Assuming that the discharge current has no electronic component,
i.e. $I_d \approx I_i$, one can write:
\begin{equation}
I_d\propto \alpha \cdot \dot{m}\propto \alpha \cdot \frac{1}{L}hd.
\label{Eq:Current}
\end{equation}
These relations between the performances and the dimensions are in
agreement with those described by Daren et al. \cite{DarenYu}, if
one considers, like they do, a constant number density $n_n$, a
constant discharge voltage $U_d$ and a geometric similarity
$R\propto r$. Using the relation \ref{Eq:Current} for the discharge
current, one finds out that the applied power $P = U_d\,I_d$ depends
on the thruster characteristic dimensions. In like manner, the
thrust efficiency is expected to depend on the thruster geometry.
The efficiency is defined as:
\begin{equation}
\eta = \frac{T^2}{2 \dot{m} U_d I_d}.
\end{equation}
The efficiency must account for the gas flow rate injected through
the anode and supplied to the cathode. Most of the time only the gas
flow injected into the channel is considered; one then talks about
the anode efficiency. Surprisingly, due to the chosen assumptions,
the anode thrust efficiency $\eta$ is solely a function of the
propellant conversion coefficient $\alpha$.

\section{Thermal constraint}
During thruster operation, a certain percentage of the input power
\emph{P} is lost due to plasma-wall interactions. Indeed, as shown
in~\cite{interaction}, a relatively large energy flux $q_p$ is
deposited onto the discharge channel walls, mostly due to ion and
electron bombardment, which results in a temperature increase of all
thrusters components. Naturally, there is maximum amount of power
that can be passed to the walls in order to limit the thermal load
and to minimize the sputtering yield of the wall material, usually
BN-SiO$_2$. On can then easily set a maximum wall temperature
$T_{max}$ over which the efficient operation of a Hall thruster is
not possible. The temperature $T_{max}$ therefore represents a
thermal constraint and it must be accounted for when designing a
thruster.

\begin{figure}[t]
\centering
\includegraphics[width=11cm]{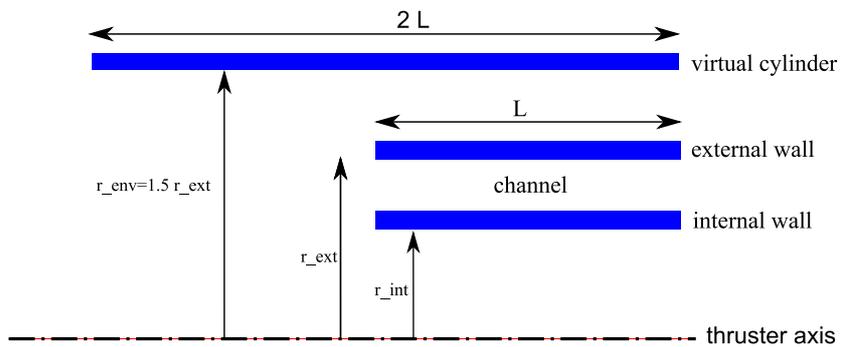}
\caption{Drawing of the simple geometry used in the channel thermal
model.} \label{fig:HET}
\end{figure}

A semi-empirical time-dependent thermal model of a Hall effect
thruster has recently been developed in order to determine the
energy flux $q_p$ from a measurement of the temporal evolution of
the channel wall temperature~\cite{interaction}. Yet, this model can
be used the other way, i.e. to determine the wall temperature from
the applied power and the channel sizes. Here, a simplified model of
the thruster discharge chamber is used. The thermal enclosure is
solely composed of the external and the internal cylindrical walls,
as can be seen in Fig.~\ref{fig:HET}, meaning that the anode and the
rear part of the channel are not taken into account. A virtual
cylinder that surrounds the enclosure is used to simulate the
channel environment (coils, magnetic core...), as explained
in~\cite{interaction}. In this work, only the steady-state wall
temperature, i.e. the equilibrium temperature, is of relevance,
meaning that the transient regime is ignored. Moreover, our previous
studies reveal that, to a great extent, heat conduction through
walls can be neglected, only taking radiation heat transfer into
account~\cite{interaction}. Under these assumptions, the total
energy leaving a grey surface \emph{$A_i$} is $\epsilon_i\sigma
T_i^4A_i$, where $\sigma$ is the Stefan-Boltzmann constant, and the
fraction of energy radiated towards another surface \emph{$A_j$} can
be calculated by multiplying the total radiated energy with the grey
body configuration factor \emph{$G_{ij}$}~\cite{interaction}. The
uniform temperature of the virtual cylinder is assumed to be
$T_{env}=450$\,K. As the temperature of the ceramic thruster walls
is much higher, the virtual cylinder can be regarded as a mere heat
sink. In order to assess the wall temperature, two quantities are
needed: the energy flux $q_p$ as well as the thruster dimensions.
The total power transferred to the channel walls $P_{loss}$ was
found to be of about 10\,\% of the input power $P$ for various Hall
thruster~\cite{interaction}. It was also demonstrated that
$P_{loss}$ is almost split in equal amounts among the internal and
external walls~\cite{interaction}:
\begin{equation}
P_{loss,int} = A_{int} \times q_{p,int} = P_{loss,ext} = A_{ext}
\times q_{p,ext} = \frac{1}{2}\,P_{loss}.
\end{equation}
Knowing the thruster dimensions $L$, $h$ and $d$, the wall
temperature $T$ can be computed as a function of the input power.
One must then check that the wall temperature does not exceed
$T_{max}$. The validity of this simple thermal model has been put to
the test by comparing the calculated and the measured wall
temperature at a given applied power for several Hall thrusters.
Results obtained with the PPSX000 thruster are displayed in
Table~\ref{TestThermalModel}. They clearly reveal a good agreement
as the gap stays below 10\,\%.\\

\noindent With BN-SiO$_2$ walls, a consistent value for $T_{max}$ is
970\,K in compliance with outcomes of a study on the thermal
behavior of Hall thrusters performed a few years
ago~\cite{MAZ_THERMO}. The maximum authorized input power that
corresponds to $T_{max}$ has been computed for a group of thrusters
with different characteristic dimensions using a power loss factor
of 10\,\%. Results are shown in Table~\ref{maxpower}. Numbers are
consistent with the range of operation power of tested thruster but
for the micro-Hall and the NASA-built thruster. In the case of the
micro-thruster, the computed maximum input power is only $9$\,W
whereas this thruster operates at a power level of $10 -
40$\,W~\cite{micro}. This is not surprising as the thruster is
equipped with a water cooling systems for the channel walls. The
estimated maximum input power for the NASA-457M thruster is only of
about $20$\,kW although this thruster is designed to be utilized
around $50$\,kW. Yet, the origin of the discrepancy is well
identified. In the thermal model, the power losses due to plasma
wall interactions are supposed to be 10\,\% of the total input power
for all thrusters without taking into account a possible size
effect. As was shown in a previous work~\cite{interaction}, this
percentage is a function of the thruster size and it decreases when
the size rises. In short, the trend originates from the fact that
the ratio of surface ($\sim$~loss term) to volume ($\sim$~thrust
generation term) decreases with the size. Indeed, when neglecting
losses at the channel back, the surface to volume ratio is equal to
1/$h$ for the channel of a HET. As a consequence, the percentage of
power losses in the NASA-475M must be below 10\,\%. The thermal
model gives a value of 5\,\% when the wall temperature $T_{max}$ is
fixed to 970~K.

Naturally, numbers given in Table~\ref{maxpower} do not represent an
upper limit in input power and a thruster can run at a power higher
than the one given by the thermal model, as the BN-SiO$_2$ ceramic
can stand a temperatures larger than $970$\,K. The efficiency as
well as the lifetime are nonetheless affected when the channel wall
temperature is in excess of 1100\,K~\cite{PPS1350}.

\section{An empirical approach to the sizing method}
\subsection{Description of the database}
A thorough open literature search using a wide range of resources
combined with data-gathering within the French research program on
electric propulsion allowed us to create a large database on Hall
effect thrusters performances. The database contains information
about thruster geometry as well as performances, notably the thrust
\emph{T}, the specific impulse \emph{$I_{sp}$} and the efficiency
\emph{$\eta$} for a series of 23 different single-stage Hall
thrusters. Moreover, the database includes information about the
magnetic field strength \emph{B}, the discharge channel wall
materials and the propellant gas. The entire database covers a vast
range of input power that stretches from 10\,W up to 50\,kW and a
large collection of data points in terms of applied discharge
voltage and gas mass flow rate. A broad range of thrust level is
certainly covered, going from 0.4\,mN with a micro Hall thruster up
to the 2.95\,N delivers by the high-power thruster developed at
NASA. The database also incorporates a few data about anode layer
thrusters (TAL) of which the distinguishing features are
a conducting channel wall and a short channel length~\cite{Zhurin}.\\

\begin{figure}[t]
\centering
\includegraphics[width=11cm]{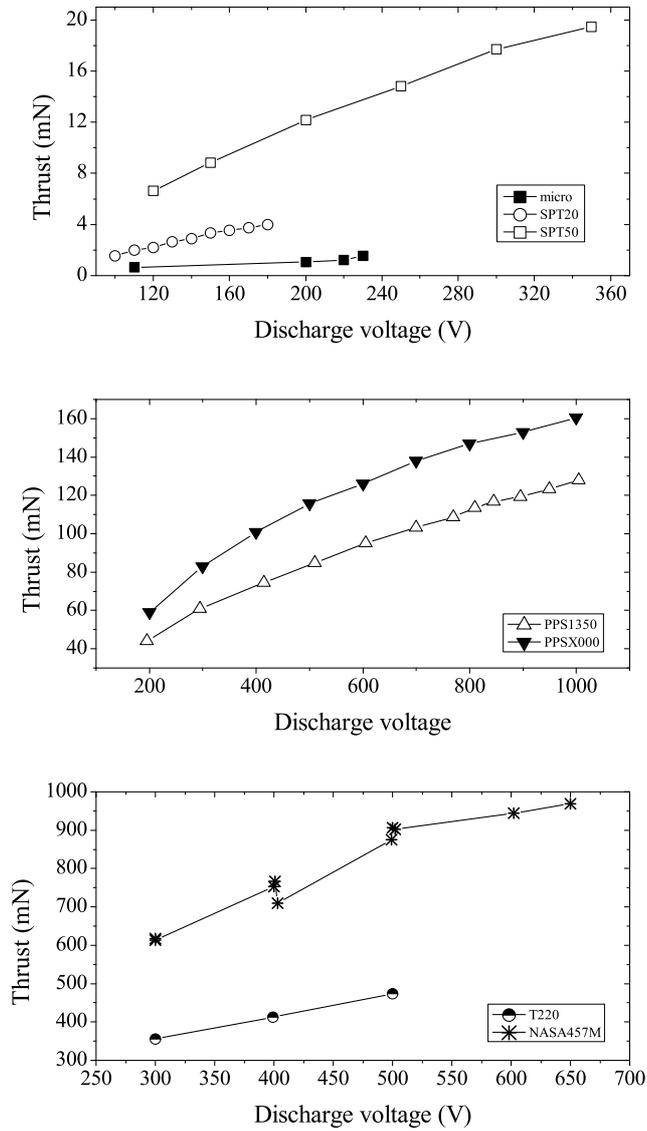}
\caption{Thrust as a function of the discharge voltage for seven
different Hall thrusters: micro-thruster (0.2\, mg/s), SPT20
(0.472\, mg/s), SPT50 (1.0\, mg/s), PPS1350 (3.5\, mg/s), PPSX000
(5.0\,mg/s), T220 (19.4\, mg/s) and NASA-457M (35.2\,mg/s).}
\label{fig:perfo1}
\end{figure}

\begin{figure}[t]
\centering
\includegraphics[width=11cm]{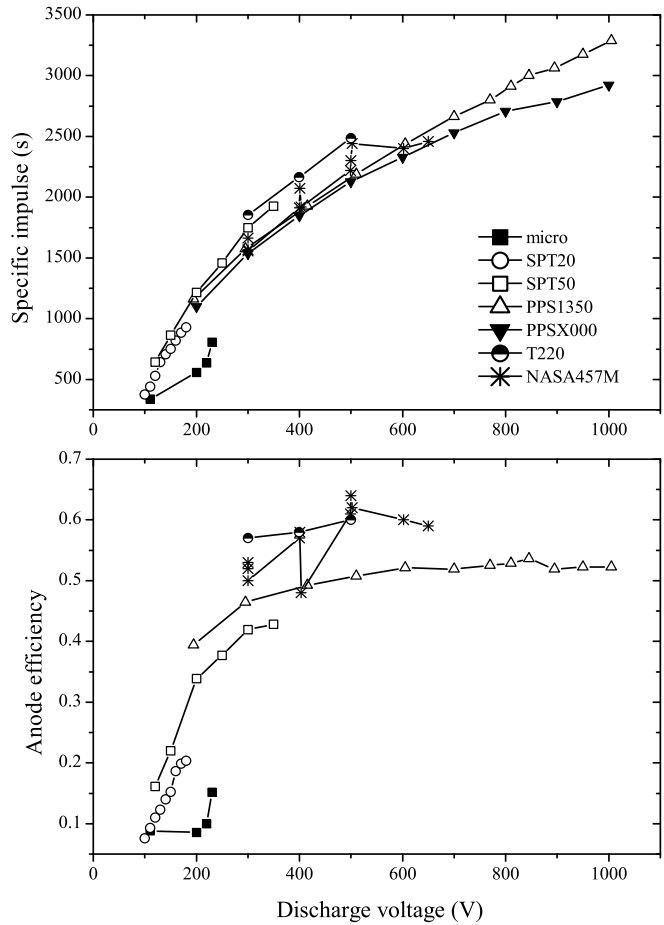}
\caption{Upper graph: Specific impulse as a function of the
discharge voltage for seven different Hall thrusters. Lower graph:
Anode thrust efficiency as a function of discharge voltage. The
xenon mass flow rate is: micro-thruster (0.2\, mg/s), SPT20 (0.472\,
mg/s), SPT50 (1.0\, mg/s), PPS1350 (3.5\, mg/s), PPSX000
(5.0\,mg/s), T220 (19.4\, mg/s) and NASA-457M (35.2\,mg/s).}
\label{fig:perfo2}
\end{figure}

A part of the collected data in terms of performance level is
displayed in Fig.~\ref{fig:perfo1} and in Fig.~\ref{fig:perfo2}. For
all thrusters, channel walls are made of BN-SiO$_2$ and the
propellant gas is xenon. The thrusters used to construct the two
figures are the following: a 4\,mm in diameter micro-Hall thruster
operating at 10-40\,W \cite{micro}, a laboratory model of the low
power SPT20 thruster~\cite{SPT20russe}, a SPT50 thruster
manufactured by the Kurchatov Institute~\cite{SPT20russe}, the
1.5\,kW-class PPS1350 HET developed and manufactured by Snecma
\cite{PPS1350}, the 5\,kW-class PPSX000 thruster which is a
laboratory version of the PPS5000 technology demonstrator developed
by Snecma \cite{PPS1350}, the 10\,kW T220 designed and built by TRW
and Space Power Inc. \cite{T220}, as well as the 50\,kW-class
NASA-457M thruster~\cite{NASA457-Xe}. The development of the thrust
as a function of the discharge voltage is shown in
Fig.~\ref{fig:perfo1} for the seven aforementioned HETs. The thrust
of course increases with $U_d$. When operating at 0.2\,mg/s and
$U_d$~=~110\,V, the micro-thruster delivers 0.4\,mN of thrust. On
the opposite side of the thrust domain, the high-power NASA-457M
thruster furnishes 970 mN of thrust when running at 35.2\,mg/s and
$U_d$~=~650\,V. The evolution of the specific impulse along with the
applied voltage is shown in Fig.~\ref{fig:perfo2}. The $I_{sp}$
increases with $U_d$ and all thrusters follow an identical trend but
the micro-thruster. The Snecma-built PPS1350 thruster delivers an
$I_{sp}$ above 3250\,s when it is fired at 1000\,V in a low gas flow
regime, as can be seen in Fig.~\ref{fig:perfo2}. An $I_{sp}$ of
3757\,s was achieved by the SPT115 thruster at 1110\,V with a gas
flow rate of 2.45\,mg/s. The behavior of the anode thrust efficiency
is shown in Fig.~\ref{fig:perfo2}. For most thrusters, a maximum is
reached around $U_d$~=~600\,V when the walls are made of BN-SiO$_2$
ceramic. This specific behavior is likely to originate from the wall
material properties combined with a change of the plasma properties
at high voltage~\cite{PPS1350}. The plot in Fig.~\ref{fig:perfo2}
reveals that the efficiency increases with the thruster size, as
discussed in the preceding section.

\subsection{Validity of the scaling laws}
\begin{figure}[t]
\centering
\includegraphics[width=10cm]{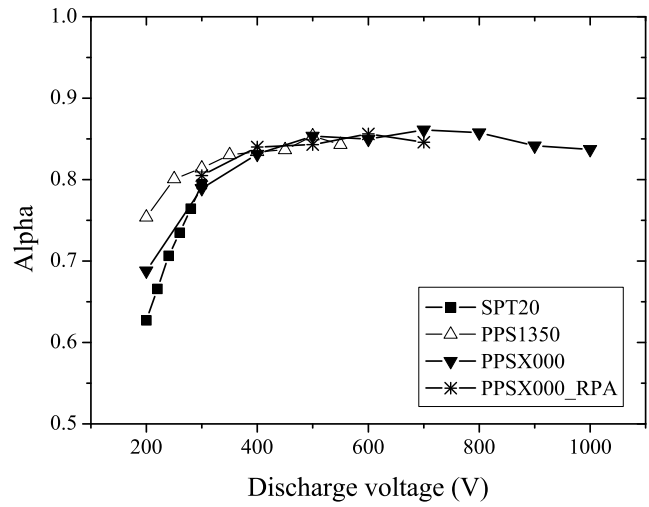}
\caption{Propellant conversion efficiency $\alpha$ as a function of
the discharge voltage for three types of Hall thrusters. The star
symbol corresponds to values obtained for the PPSX000 thruster when
the ion velocity is recorded by means of a RPA.} \label{fig:alpha}
\end{figure}

The propellant conversion efficiency \emph{$\alpha$} is the ratio of
the ion mass flow rate to the propellant mass flow rate:
\begin{equation}
\alpha=\frac{\dot{m_i}}{\dot{m}}\approx \frac{I_i}{I_m}
\end{equation}
Therefore \emph{$\alpha$} is not constant, as was considered in
section~\ref{section_relation}, but it is a function of the
discharge voltage and the mass flow rate. The value of
\emph{$\alpha$} can be obtained from the performance data using the
following equation:
\begin{eqnarray}
\label{alpha_thrust}
\nonumber T=\dot{m_i}\,\bar{v_i}=\alpha \,
\dot{m} \,
\bar{v_i}=\alpha \, \dot{m} \, \sqrt{\frac{2eU_d}{m_i}}, \\
\Rightarrow \alpha=\frac{T}{\dot{m}} \, \sqrt{\frac{m_i}{2eU_d}}.
\end{eqnarray}
Figure~\ref{fig:alpha} shows the calculated values of $\alpha$ for
three different thrusters as a function of the applied voltage $U_d$
when only Xe$^+$ ions are taken into account. This figure indicates
that $\alpha$ depends both on the thruster size and on the value of
$U_d$. The small SPT20 thruster indeed exhibits the lowest values of
$\alpha$. There is a natural link between $\eta$, $\alpha$ and the
dimensions. As can be seen in Fig.~\ref{fig:alpha}, $\alpha$
increases quickly with the applied voltage, and for a large thruster
it approaches unity at high voltage. The growth of $\alpha$ is
especially connected with the electron temperature that increases
with the voltage~\cite{RAITSES}. Actually, the production of
multiply-charged ions must also be dealt for accurately assessing
the value of $\alpha$~\cite{ALEXEY}. The growth of $\alpha$ is
connected with both the the electron temperature and the fraction of
multiply-charged ions~\cite{ALEXEY}. The two quantities increase
with the voltage. According to the collected data set, the
calculated values of $\alpha$ vary between 0.3 and 0.96, not taking
into account the micro thruster. For low voltages, $\alpha$ drops
quickly due to a weak electron temperature. For an input power
higher than 1\,kW and an applied voltage above 300\,V, the quantity
$\alpha$ is commonly in the range 0.8\,-\,0.9. The value determined
by means of Eq.~\ref{alpha_thrust} are slightly underestimated as
voltage losses are not accounted for. Nevertheless, the value of
$\alpha$ obtained when using the ion velocity measured by way of a
repulsing potential analyzer in the thruster near field are close to
the ones computed with $e \cdot U_d$ as kinetic
energy~\cite{MAZOUFFRE_RPA}. That means the voltage losses term
$\Delta$ is small.

\begin{figure}[t]
\centering
\includegraphics[width=10cm]{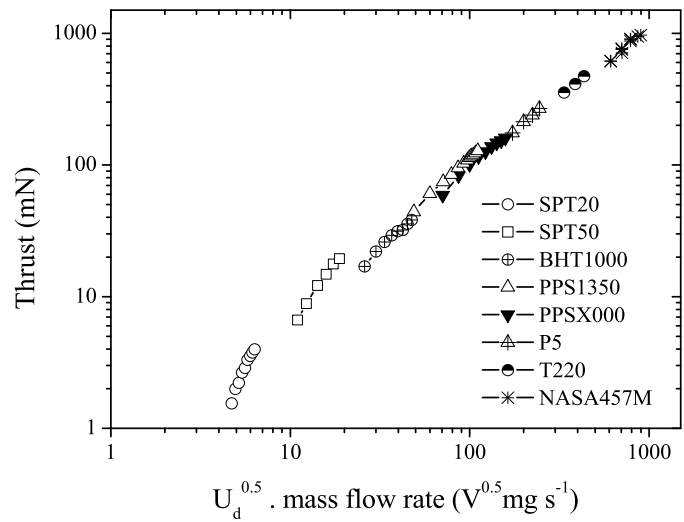}
\caption{Thrust as a function of $\dot{m}\sqrt{U_d}$ for a wide
range of thruster sizes.} \label{fig:poussee2}
\end{figure}

\begin{figure}[t]
\centering
\includegraphics[width=9cm]{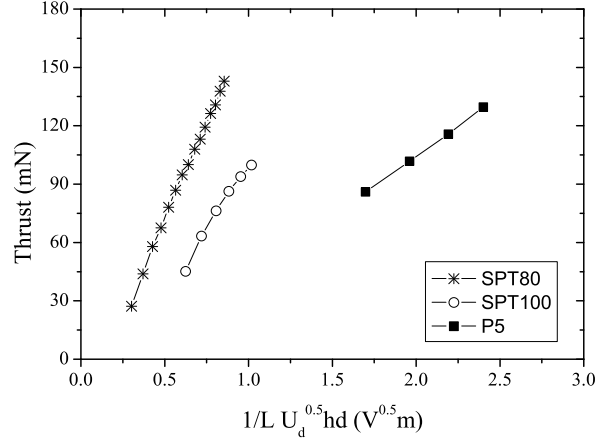}
\caption{Thrust as a function of the product
$\frac{1}{L}\sqrt{U_d}hd$ for the SPT80 at $\dot{m}=4.58$\,mg/s as
well as the SPT100 and the P5 at $\dot{m}=5$\,mg/s.}
\label{fig:poussee}
\end{figure}

\begin{figure}[t]
\centering
\includegraphics[width=9cm]{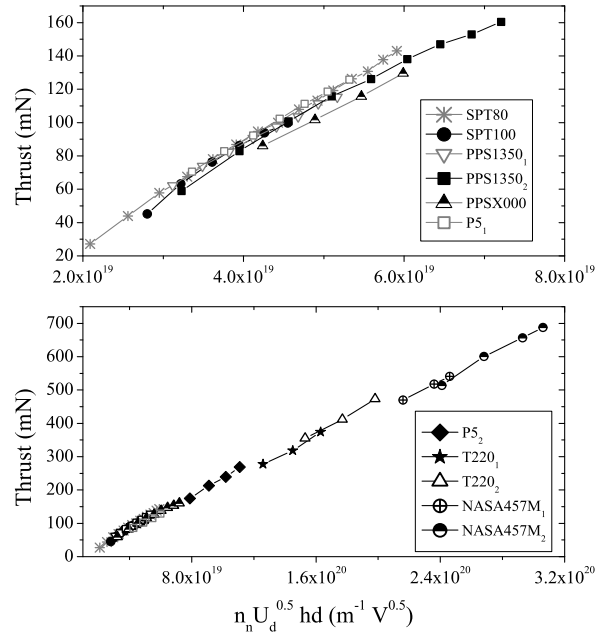}
\caption{Thrust as a function of $n_n\sqrt{U_d}hd$ for different
thrusters; Upper Graph: SPT80 (4.58\,mg/s), SPT100, PPS1350$_1$
(4.82\,mg/s), PPS1350$_2$ (5.21\,mg/s), PPSX000 and P5$_1$ at a flow
rate of about 5\,mg/s; Lower Graph: SPT80 (4.58\,mg/s), SPT100
(5\,mg/s), PPS1350$_1$ (4.82\,mg/s), PPS1350$_2$ (5.21\,mg/s),
PPSX000 (5\,mg/s), P5$_1$ (5\,mg/s), P5$_2$ (10\,mg/s), T220$_1$
(15.9\,mg/s), T220$_2$ (19.4\,mg/s), NASA-457M$_1$ (20.0\,mg/s) and
NASA-457M$_2$ (24.9\,mg/s).} \label{fig:MM}
\end{figure}

Figure~\ref{fig:poussee2} shows the thrust as the function of the
product $\dot{m}\sqrt{U_d}$ for a large collection of HETs and
operating conditions. Apart from SPT20 thruster points at low power,
all data points are aligned. As suggested earlier, the coefficient
$\alpha$ is almost constant. From the slope of the curve in
Fig.~\ref{fig:poussee2}, one finds $\alpha$~=~(0.89$\pm$0.01). Note
that data about the SPT20 thruster are not considered for the linear
fit. Equation~\ref{thrust} gives a relation between the thrust, the
voltage and the characteristic dimensions. In
Fig.~\ref{fig:poussee}, the thrust is plotted as a function of the
product $\frac{1}{L}\sqrt{U_d}hd$, as established by
Eq.~\ref{thrust}, for a variety of Hall thrusters firing with
$\dot{m} \approx~5$\,mg/s. The normal operating point of the
thrusters ranges from 500\,W to 5\,kW. As can be seen, curves are
linear, however, the slope depends upon the input power. A detailed
analysis carried out with all thrusters available in the database
reveals that thrusters group together depending on their normal
operating power. Equation~\ref{thrust} was derived using the
relation between $\dot{m}$ and the dimensions as well as the series
of assumptions listed at the beginning of section~2.1. Though graph
of Fig.~\ref{fig:poussee} indicates that some of the assumptions may
be too limiting. First, the propellant gas temperature is said to be
constant an independent of $U_d$ in such a way that the propellant
speed inside the channel is fixed and it does not vary with $U_d$.
In effect the gas temperature depends on the thruster geometry as
well as on the input power. Second, the propellant conversion
efficiency $\alpha$, which was assumed to be constant, is a function
of the discharge voltage as demonstrated hereinbefore. These two
points are confirmed by the fact that the curves for thrusters of
about the same sizes, e.g. the P5 and the PPSX000 as well as the
PPS1350 and the SPT100, are close to each others. Yet, when the
thrust $T$ is plotted as a function of the product
$n_n\,h\,d\,\sqrt{U_d}$, all datapoints are aligned, as can be
observed in Fig.~\ref{fig:MM}. The plot is independent of the power
level and the thruster sizes. This experimental fact reveals that
the Melikov-Morozov criterion does not implies that the ratio
$\lambda_i / L$ is identical for all thrusters, whatever the
geometry and design. It is even the opposite, and the mean value of
the parameter $\beta$ given in Table~\ref{TableCoef} must seen as
the typical value that warrants a fine thruster functioning.

As explained in section~\ref{section-ScalingLaws}, the input power
$P$ naturally depends upon the characteristic thruster dimensions.
Assuming $I_d \approx I_i$, $P$ is found to be a function of the
ratio $hd/L$, see Eq.~\ref{Eq:Current}. However, using all gathered
data, it appears that $P$ varies linearly with the $hd$ product, as
shown in Fig.~\ref{fig:puissance}. This linear relation will be used
in the remainder of the paper when sizing high-power Hall thrusters.

\begin{figure}[t]
\centering
\includegraphics[width=9cm]{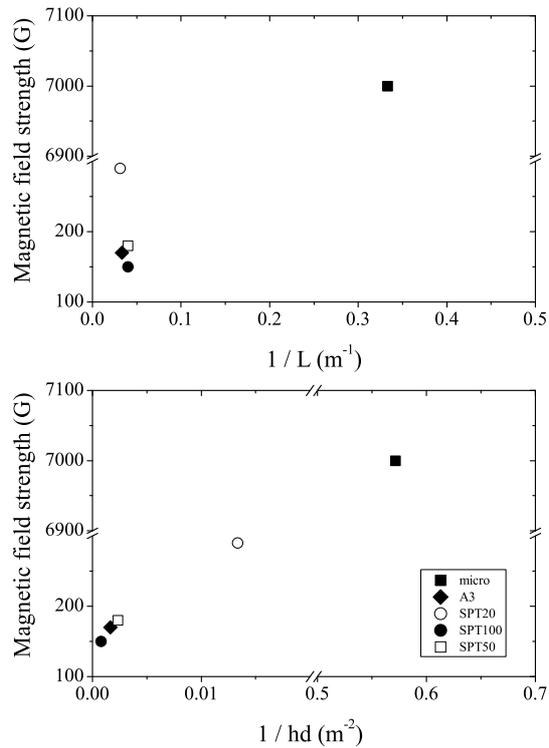}
\caption{Upper graph: Magnetic field strength as a function of
$1/L$. Lower graph: Magnetic field strength as a function of
$1/hd$.} \label{fig:B}
\end{figure}

At the end of section~\ref{section-ScalingLaws}, two scaling laws
have been established for the magnetic field $B$, see Eq.~\ref{B1}
and \ref{B2}. The two plots in Fig.~\ref{fig:B} show the evolution
of $B$ with $1/L$ and $1/hd$, respectively, for five different Hall
effect thrusters. As can be seen, data points are well aligned only
when the $hd$ product is considered. Therefore, one can consider
that the scaling laws $B \propto 1/hd$ is more suitable than the one
that solely incorporates the channel length $L$, although it was
never considered up until then.

Finally, a last remark is worth making. The atom density $n_n$ that
warrants an efficient functioning of a Hall effect thruster can be
inferred from the xenon mass flow rate when the thruster dimensions
are known, see Eq.~\ref{debit}. Assuming a propellant temperature of
800\,K inside the channel and taking the gas mass flow rate at
normal operating conditions, one finds $n_n \approx
1.2\times10^{19}$\,m$^{-3}$ whatever the thruster picked in the
database, but the micro-thruster. This number, which can be
envisaged as an \emph{atom density constraint}, is in good agreement
with the value commonly found in literature.

\subsection{Determination of the proportionality coefficients}
In order to assess the required thruster dimensions by way of a
scaling method for an available input power and a given thrust
level, it is necessary to know the proportionality coefficients of
equations encountered in section~2.2. As it was previously shown,
the specific impulse is proportional to $\sqrt{U_d}$:
\begin{eqnarray}
\nonumber & & I_{sp}=C_{I_{sp}}\cdot \sqrt{U_d}\\
 & & \mbox{with}\quad C_{I_{sp}}=\frac{\sqrt{\frac{2e}{m_i}}}{g_0}=123.4
\label{constIsp}
\end{eqnarray}
This coefficient can also be inferred from the experimental curve
$I_{sp}$ as a function of $\sqrt{U_d}$, see Fig.~\ref{fig:perfo2}.
The mean value of the coefficient $C_{I_{sp}}^*$ determined by using
the database is \(C_{I_{sp}}^*=99.7\).

The thrust is described by the following equation:
\begin{eqnarray}
\nonumber & & T=C_{T_1}\cdot \dot{m}\,\sqrt{U_d}\\
 & & \mbox{with} \quad C_{T_1}=\alpha \sqrt{\frac{2e}{m_i}}=1090.8,
\label{thrust-mass flow}
\end{eqnarray}
when using $\alpha = 0.9$. The coefficient $C_{T_1}$ is the slope of
the curve \emph{T} versus $\dot{m}\sqrt{U_d}$, see
Fig.~\ref{fig:poussee2}. Using the experimental dataset, one obtains
\(C_{T_1}^*=1077.3\). A second relation was established for the
thrust:
\begin{eqnarray}
\nonumber & & T=C_{T_2} \cdot \frac{1}{L} \, \sqrt{U_d} \, hd\\
 & & \mbox{with} \quad C_{T_2}=\alpha \, m_n \, v_n \, \pi \,
\sqrt{\frac{2e}{m_i}} \, \frac{v_n}{\beta \sigma_i v_e}
\label{thrust-dimension}
\end{eqnarray}
Assuming that the electron temperature is 10\,eV and the gas
temperature is 800\,K, the corresponding velocities are
$v_e=2\times10^6$\,m/s and $v_n=320$\,m/s, respectively.
Furthermore, $n_n$ is taken to be $10^{19}\,{\rm m}^{-3}$ and
$\sigma_i$ is equal to $5\times10^{-20}\,{\rm m}^2$. Using those
numbers, one finds $C_{T_2}=7.65 \times 10^{-4}/\beta$. The value of
the $\beta$ coefficient can be determined using our database as the
size $L$ is known for all thrusters. With numbers given above, one
finds $\beta$~=~0.007, and it comes $C_{T_2}=0.109$. The true
experimental value $C_{T_2}^*$ is obtained by way of the type of
curves plotted in Fig.~\ref{fig:poussee}, notwithstanding the fact
that the slope of relation~\ref{thrust} depends upon the thruster
normal operating power as discussed previously. A mean value is
$C_{T_2}^*$~=0.0924.

\begin{figure}[t]
\centering
\includegraphics[width=9cm]{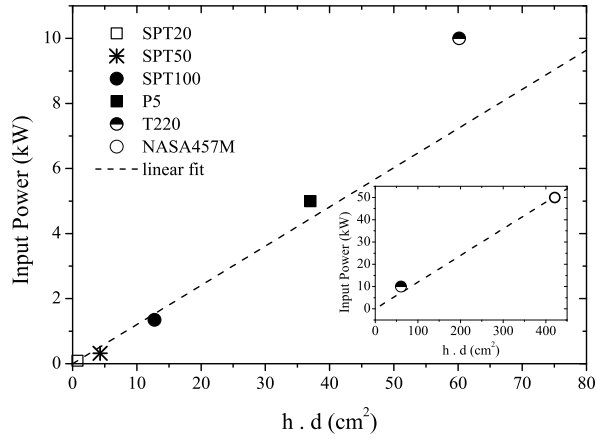}
\caption{Input power as a function of the product $h \times d$ for
six Hall thrusters. Also shown is a linear fit through all data
points.} \label{fig:puissance}
\end{figure}

As shown in Fig.~\ref{fig:puissance}, the input power is a function
of the product $hd$ under our assumptions, and one can write:
\begin{eqnarray}
\nonumber & & P = C_P \cdot hd. \\
 & & \mbox{with} \quad C_P = \frac{e\,U_d\,\alpha\,\dot{m}}{m_n\,hd} \approx 1.1\times10^{6}.
\label{PCPrelation}
\end{eqnarray}
The value of $C_p$ is calculated from the collection of operating
points selecting various thrusters to vary the $hd$ product. The
empirical value of the proportionality coefficient is taken to be
the slope of the line in Fig.~\ref{fig:puissance}: $C_P^* =
1.2\times10^{6}$. The value of all coefficients are summarized in
Table~\ref{TableCoef}. Finally, the thruster dimensions can be
assessed afterwards, as it will be shown in the next section.

\section{Design of a high power Hall Effect Thruster}
\label{section:design}

High-power Hall effect thrusters in the range 10\,-\,30\,kW and able
to deliver a thrust level around 1\,N with a specific impulse of
about 2500\,s are thought to be used as the primary propulsion
system for robotic space probes during interplanetary
journeys~\cite{JPL,HP1,HP2}. Such high-power Hall thrusters may also
be of interest for orbit transfer maneuvers of large satellites.
Only a few high-power prototypes have been developed in the world so
far and a significant research effort on this specific technology is
now appearing within Europe. For these reasons, the sizing method
based on aforementioned widely applicable scaling laws in
combination with our large database is employed to design a
20\,kW-class thruster with $T=1$\,N as well as a 25\,kW-class Hall
thruster with $T=1.5$\,N.

The discharge voltage of the two thrusters is fixed to $U_d=500$\,V,
i.e. $I_{sp}$~=~2760\,s when considering singly-charged ions and no
voltage losses. Xenon is used as a propellant gas. The channel walls
are assumed to be made of BN-SiO$_2$ ceramics. The highest wall
temperature is set to 970\,K. In this study, losses are fixed to
6\,\% of the applied power. The parameters $\alpha$ and $\beta$ are
set to 0.9 and 0.007, respectively, in agreement with previous
analysis.

\subsection{Sizing of a 20\,kW-class thruster}
The process that consists in determining the thruster characteristic
dimensions $L$, $h$ and $d$ as well as the magnetic field $B$ must
be carried out step by step.
\begin{enumerate}
\item The required mass flow rate is determined by means of Eq.~\ref{thrust-mass
flow}.
\item The $hd$ product is found using the
relationship~\ref{PCPrelation}.
\item The channel length $L$ is given by Eq~\ref{thrust-dimension}.
\item The thermal constraint is used to assess the value of both the
channel width $h$ and the average channel diameter $d$. The process
can be iterative, in the case the dimensions must be changed in
order to satisfy the constraint: wall temperature below $T_{max}$.
\item The magnetic field strength is then obtained from the relation $B \propto 1/hd$, see
Fig.~\ref{fig:B}.
\item At last, it should be verified that the number
density $n_n$ is close to $1.2\times10^{19}$, as explained at the
end of section~4.2.
\end{enumerate}
The channel length is the dimension that is the most difficult to
determine as the proportionality coefficient $C_{T_2}$ strongly
depends on the value of $\beta$. In fact, the ratio between
$\lambda_i$ and $L$ is not constant but it can vary considerably for
different mass flow rates, voltages and thruster types. Since the
value of the $hd$ product given by the method is quite reliable, it
appears better to modify rather the channel length than the mean
diameter in case an iterative loop is necessary to satisfy the
thermal constraint. Indeed, outcomes of the thermal model are not
much influenced by a change in the value of $h$ as losses are
fixed.\\

For the 20\,kW Hall thruster (1\,N, 500\,V), the sizing method leads
to: $\dot{m}=41.1$\,mg/s, $L=40.1$\,mm, $d=250$\,mm and $h=66$\,mm
and a magnetic field strength $B=136$\,G. The temperature is
$T_{ext}=865$\,K and $T_{int}=962$\,K for the external, respectively
the internal, dielectric channel wall. Finally, the value
$1.12\times10^{19}$\,m$^{-3}$ is found for the atom number density.

\subsection{Sizing of a 25\,kW-class thruster}
The same approach was applied to design a 25\,kW Hall thruster
(1.5\,N, 500\,V). The sizing method gives: $\dot{m}=61.6$\,mg/s,
$L=40$\,mm, $d=290$\,mm and $h=71$\,mm. The magnetic field strength
$B$ is equal to 135\,G. The temperature is $T_{ext}=871$\,K and
$T_{int}=979$\,K for the external, respectively the internal,
dielectric channel wall. The calculated number density is
$n_n=1.35\times 10^{19}$\,m$^{-3}$.

The outcomes of our sizing method can be compared with results
obtained in a study by M.~Andrenucci and co-workers in which sizes
are fixed instead of being computed~\cite{Andrenucci}. They used
their specific approach to design a 25\,kW Hall thruster operating
at 275\,V. The dimensions are fixed to $L=30$\,mm, $d=250$\,mm and
$h=62$\,mm. They found a xenon mass flow rate of 88.7\,mg/s, a
thrust level of 1.49\,N and a magnetic field of 135\,G. Our
procedure leads to a mass flow rate of 83.4\,mg/s, an atom density
of $2.61\times10^{19}$\,m$^{-3}$ and the same magnitude for $B$.
Moreover, their thruster dimensions do not permit to respect our
thermal constraint as $T_{ext}=970$\,K and $T_{int}=1103$\,K.
Nevertheless, to achieve a thrust level of 1.49\,N with
$U_d$~=~275\,V, our approach predicts $L=35$\,mm, $d=310$\,mm and
$h=67$\,mm, values that are not too far from the ones chosen a
priori referring to a vast database~\cite{Andrenucci}.

\section{Conclusion}

The Hall effect thruster sizing method described in these works
considers the three characteristic thruster dimensions $L$, $d$ and
$h$, as well as the magnetic field strength $B$. The method relies
on analytical laws that are established from the fundamental
principles that govern the physics of a Hall thruster in the frame
of simplifying assumptions. Besides, the thruster geometry must
fulfill criteria about channel wall temperature and atomic number
density. A vast database that incorporates 23 single-stage Hall
thrusters covering a power range between 10\,W and 50\,kW allows to
check the validity of scaling laws and to find the value of
corresponding coefficients necessary to proportion a thruster. The
sizing approach was then employed to obtain a proper estimate of
characteristic dimensions and magnetic field strength for a 20\,kW
and a 25\,kW Hall thruster capable of providing a thrust level of
1\,N, respectively 1.5\,N. Our approach gave also satisfactory
results when it was applied to check the design of a 100\,W Hall
thruster currently under development and test in our laboratory.

Scaling laws developed here solely represent a first order approach
due to the numerous simplifying assumptions on the physics at work
in a Hall thruster. Nevertheless, for a given set of operating
conditions they furnish a first estimate of the geometry and the
magnetic field strength of a thruster, which permits to save time
during the design and optimization stages. One way to improve the
accuracy of the scaling method is to reduce the number of
assumptions. The evolution of the electron temperature, of the gas
temperature and of the fraction of multiply-charged ion species as a
function of the discharge voltage could for instance be taken into
account, however, available data are scarce that anyway limits the
gain. Another way could consist in applying a statistical analysis
of the vast database to directly obtain empirical sizing formulas.
The design of experiment method could be an appropriate tool to
achieve that goal. Finally, attempt to incorporate the magnetic
field topology into scaling laws would represent a tremendous
progress as the latter is the most fundamental feature to ensure a
successful operation.

\section*{Acknowledgments}
This work is carried out in the frame of the
CNRS/CNES/SNECMA/Universit\'es joint research program 3161 entitled
``\emph{Propulsion par plasma dans l'espace}''.

\newpage
\begin{table}
\caption{Comparison between the measured PPSX000 thruster external
and internal wall temperatures and the ones obtained with a simple
thermal model for three values of $P$. The level of power losses is
fixed to 10\,\%.} \centering \small
\begin{tabular}{ccccc}
\hline\hline
Power (W) & \multicolumn{2}{c}{$T_{ext}$ (K)} & \multicolumn{2}{c}{$T_{int}$ (K)} \\
 & Mes. & Calc. & Mes. & Calc. \\
\hline
1605 & 690 & 655 & 740 & 710 \\
2460 & 762 & 729 & 788 & 790 \\
4620 & 833 & 854 & 862 & 926 \\
\hline\hline
\end{tabular}
\label{TestThermalModel}
\end{table}

\newpage
\begin{table}
\caption{Outcomes of the thermal model in terms of maximum input
power for various thruster types when $T_{max}$ is set to 900\,K.
The level of power losses is fixed to 10\,\%.} \centering \small
\begin{tabular}{lc} \hline\hline
Thruster & Maximum input \\
 & power (W)\\ \hline
micro & 9 \\ 
SPT20 & 210 \\ 
SPT50 & 870 \\ 
PPS1350 & 2060 \\ 
SPT100 & 2080 \\ 
PPSX000 & 4200 \\ 
P5 & 5550 \\ 
NASA-457M & 20300 \\ \hline\hline
\end{tabular}
\label{maxpower}
\end{table}

\newpage
\begin{table}
\caption{Values of all proportionality coefficients needed for
scaling laws. Values are in standard units.} \centering \small
\begin{tabular}{cc}
\hline\hline
Coefficient & Value\\
\hline
$C_{I_{sp}}$ & 123.4\\
$C_{I_{sp}}^*$ & 99.7\\
$C_{T_{1}}$ & 1090.8\\
$C_{T_{1}}^*$ & 1077.3\\
$C_{T_{2}}$ & 0.109\\
$C_{T_{2}}^*$ & 0.092\\
$C_P$ & $1.1\times10^6$\\
$C_P^*$ & $1.2\times10^6$\\
$\beta$ & 0.007\\
\hline\hline
\end{tabular}
\label{TableCoef}
\end{table}


\end{document}